\documentclass[12pt]{article}
\usepackage{wider}
\usepackage{graphicx}

\begin{document}

\newcommand{\sqvb}{\ensuremath{ \langle \!\langle 0 |} }
\newcommand{\sqvk}{\ensuremath{ | 0 \rangle \!\rangle } }
\newcommand{\sqvn}{\ensuremath{ \langle \! \langle 0 |  0 \rangle \! \rangle} }
\newcommand{\wh}{\ensuremath{\widehat}}
\newcommand{\be}{\begin{equation}}
\newcommand{\ee}{\end{equation}}
\newcommand{\bea}{\begin{eqnarray}}
\newcommand{\eea}{\end{eqnarray}}
\newcommand{\ra}{\ensuremath{\rangle}} 
\newcommand{\la}{\ensuremath{\langle}}
\newcommand{\rra}{\ensuremath{ \rangle \! \rangle }}
\newcommand{\lla}{\ensuremath{ \langle \! \langle }}
\newcommand{\str}{\rule[-.125cm]{0cm}{.5cm}}
\newcommand{\pr}{\ensuremath{^\prime}}
\newcommand{\ppr}{\ensuremath{^{\prime \prime}}}
\newcommand{\da}{\ensuremath{^\dag}}
\newcommand{\as}{^\ast}
\newcommand{\eps}{\ensuremath{\epsilon}}
\newcommand{\ve}{\ensuremath{\vec}}
\newcommand{\ka}{\kappa}
\newcommand{\non}{\ensuremath{\nonumber}}
\newcommand{\lf}{\ensuremath{\left}}
\newcommand{\rt}{\ensuremath{\right}}
\newcommand{\al}{\ensuremath{\alpha}}
\newcommand{\dfn}{\ensuremath{\equiv}}
\newcommand{\ga}{\ensuremath{\gamma}}
\newcommand{\ti}{\ensuremath{\tilde}}
\newcommand{\wti}{\ensuremath{\widetilde}}
\newcommand{\hs}{\ensuremath{\hspace*{.5cm}}}
\newcommand{\bet}{\ensuremath{\beta}}
\newcommand{\om}{\ensuremath{\omega}}

\newcommand{\cO}{\ensuremath{{\cal O}}}
\newcommand{\cS}{\ensuremath{{\cal S}}}
\newcommand{\cF}{\ensuremath{{\cal F}}}
\newcommand{\cX}{\ensuremath{{\cal X}}}
\newcommand{\cZ}{\ensuremath{{\cal Z}}}
\newcommand{\cG}{\ensuremath{{\cal G}}}
\newcommand{\cR}{\ensuremath{{\cal R}}}
\newcommand{\cV}{\ensuremath{{\cal V}}}
\newcommand{\cC}{\ensuremath{{\cal C}}}
\newcommand{\cP}{\ensuremath{{\cal P}}}
\newcommand{\pup}{\ensuremath{^{(p)}}}
\newcommand{\prpr}{\ensuremath{\prime \prime }}

\newcommand{\ltsimeq}{\raisebox{-0.6ex}{$\,\stackrel
        {\raisebox{-.2ex}{$\textstyle <$}}{\sim}\,$}}
\newcommand{\gtsimeq}{\raisebox{-0.6ex}{$\,\stackrel
        {\raisebox{-.2ex}{$\textstyle >$}}{\sim}\,$}}
\newcommand{\prpsimeq}{\raisebox{-0.6ex}{$\,\stackrel
        {\raisebox{-.2ex}{$\textstyle \propto $}}{\sim}\,$}}

%To see labels in DVI:
%\newcommand{\xxx}[1]{\hspace*{\fill}{\bf #1}\\}
%\newcommand{\yyy}[1]{\hspace*{\fill}{\bf #1}\\}
%\newcommand{\zzz}{ {\bf HAND CITATION} }
%NB No underscore allowed in name.
%To hide labels in DVI:
\newcommand{\xxx}[1]{}
\newcommand{\yyy}[1]{}
\newcommand{\zzz}{}

\title{\bf 
Loss-Induced Limits to Phase Measurement Precision with Maximally Entangled States
\footnote{This work was sponsored by the Air Force under Air Force Contract 
FA8721-05-C-0002.  Opinions, interpretations, conclusions, and 
recommendations are those of the authors and are not necessarily endorsed 
by the U.S. Government.}
}
\author{Mark A. Rubin and Sumanth Kaushik\\
Lincoln Laboratory\\ 
Massachusetts Institute of Technology\\  
244 Wood Street\\                         
Lexington, Massachusetts 02420-9185\\%}
\{rubin,skaushik\}@LL.mit.edu\\
}
\date{\mbox{}}
\maketitle

\begin{center}
\begin{abstract}
The presence of loss limits the precision of an approach to phase measurement
using maximally entangled states, also referred to as NOON  states.  
A calculation using a simple beam-splitter model
of loss shows that, for all nonzero values $L$\/ of the loss, phase measurement precision degrades with increasing
number $N$\/ of entangled photons  for $N$\/ sufficiently large.
For  $L$\/  above a critical value of approximately 0.785, phase measurement precision degrades with
increasing $N$\/ for all values of $N$\/.  
For  $L$\/ near  zero, phase measurement precision  improves with
increasing $N$\/ down to a 
 limiting precision of approximately $1.018 L$\/ radians, 
attained at $N$\/ approximately  equal to $2.218/L$\/, and degrades as $N$\/ increases beyond this value.
Phase measurement precision with multiple measurements and a fixed total
number of photons $N_T$\/ is  also examined. For $L$\/ above a critical
value of approximately 0.586, the ratio of phase measurement precision attainable with
NOON states to that attainable by conventional methods using unentangled coherent states degrades with
increasing $N$\/, the number of entangled photons employed in a single measurement, 
for all values of $N$\/.  For $L$\/ near  zero this ratio is optimized by using approximately 
$N=1.279/L$\/ entangled photons in each
measurement, yielding a precision of approximately $1.340\sqrt{L/N_T}$\/ radians.
\end{abstract}
\end{center}

\section{NOON States and the Heisenberg Limit}

The use of entangled states has been proposed 
\cite{Yurke1986}-\cite{Leeetal2004}% {Camposetal2003}
as a means of performing phase measurements 
with a precision $\delta \phi_{min}$\/ at the Heisenberg limit. In this limit, $\delta \phi_{min}$\/ scales as
\be
\delta \phi_{min} \sim 1/N,\label{Heisenberglimit}
\ee
\xxx{Heisenberglimit}
with increasing photon number $N$\/, rather than  at the standard quantum limit
\be
\delta \phi_{min} \sim 1/\sqrt{N}.\label{SQL}
\ee
\xxx{SQL}
Entangled-state enhancements to related tasks such as frequency measurement and lithography
have also been proposed   \cite{Winelandetal1992}-\cite{KapaleDowling2006}.
%\cite{Giovannettietal2006}.
Experiments implementing phase measurements  and related tasks using entangled states have been performed 
for the cases of $N=2$\/ \cite{Ouetal1990a}-\cite{Benninketal2004}, $N=3$\/ \cite{MitchellLundeenSteinberg2004}
and $N=4$\/ \cite{Waltheretal2004}.

Maximally entangled states, also referred to as NOON states \cite{GerryKnight2005}, are states of the form
\be
|N::0\ra_{a,b}=\frac{1}{\sqrt{2}}\left(|N,0\ra_{a,b}+|0,N\ra_{a,b}\right),\label{NOONstatedef}
\ee
\xxx{NOONstatedef}
\yyy{p. NSL-1 box1}
where
\be
|m,n\ra_{a,b}=|m\ra_a|n\ra_b,\label{prodstatedef}
\ee
\xxx{prodstatedef}
and where $|m\ra_a$\/ is a Fock state with $m$\/ quanta in mode $a$\/,
\be
|m\ra_a=\frac{1}{\sqrt{m!}}\left({\wh{a}_a}\da\right)^m|0\ra_a,\label{numberstatedef}
\ee
\xxx{numberstatedef}
with ${\wh{a}_a}\da$\/ and  $|0\ra_a$\/  the usual creation operator and vacuum state  for mode $a$\/.
In interferometry, for example,
modes $a$\/ and $b$\/ are different paths around the interferometer. 
The argument that NOON states allow phase measurement at the Heisenberg limit is as follows. 

A phase shift of $\phi$\/ in
mode $b$\/ changes the state (\ref{NOONstatedef}) to
\be
|N::0;\phi\ra_{a,b}=\frac{1}{\sqrt{2}}\left(|N,0\ra_{a,b}+\exp(iN \phi)|0,N\ra_{a,b}\right).\label{NOONphistatedef}
\ee
\xxx{NOONphistatedef}
\yyy{p. NSL-1 box2}
The phase $\phi$\/ can be determined by  measuring the operator \cite{Koketal2002,Leeetal2002,MitchellLundeenSteinberg2004}
\be
\wh{A}_N=|0,N\ra_{a,b}\;\;\mbox{}_{a,b}\la N,0|+|N,0\ra_{a,b}\;\;\mbox{}_{a,b}\la 0,N|.\label{ANdef}
\ee
\xxx{ANdef}
\yyy{p. NSL-1 box 3}
In the state (\ref{NOONphistatedef}), the expectation value of the operator (\ref{ANdef}) is
\bea
\la \wh{A}_N \ra_{\phi}&=&\mbox{}_{a,b}\la N::0 \; ;\phi |\wh{A}_N |N::0 \; ;\phi \ra_{a,b}\nonumber\\ 
                       &=&\cos(N\phi), \label{meanANphi}
\eea
\xxx{meanANphi}
\yyy{p. NSL-1 last box}
and its variance is
\bea
\mbox{Var}_{\phi}\wh{A}_N&=&\mbox{}_{a,b}\la N::0 \; ;\phi |\wh{A}_N ^2|N::0 \; ;\phi\ra_{a,b}-
\left(\mbox{}_{a,b}\la N::0  \; ;\phi |\wh{A}_N |N::0  \; ;\phi\ra_{a,b}\right)^2\nonumber\\
                         &=&\sin^2(N\phi).\label{VarANphi}
\eea
\xxx{VarANphi}
\yyy{p. NSL-2 box}
The signal-to-noise ratio (SNR) for detecting a change $\delta \phi$\/ about a phase value $\phi_0$\/
is\cite{Helstrom1976}
\be \mbox{SNR}=\left(\la \wh{A}_N\ra_{\phi_0+\delta \phi} -\la \wh{A}_N\ra_{\phi_0}\right)^2/\mbox{Var}_{\phi_0}\wh{A}_N.
\label{SNRdef}
\ee
\xxx{SNRdef}
\yyy{p. NSL-2 box 8}
Using (\ref{meanANphi}) and (\ref{VarANphi}) in (\ref{SNRdef}), 
\be
\mbox{SNR}=N^2\left(\delta \phi \right)^2\label{SNRnoloss}
\ee
\xxx{SNRnoloss}
\yyy{p. NSL-2 last box}
for small phase changes,
\be
|\delta \phi| \ll 2\pi\/. \label{smallphasechange}
\ee
\xxx{smallphasechange}
Defining the minimum detectable phase change  $\delta \phi_{min}$\/ to be that phase 
change corresponding
to an SNR of unity\cite{WallsMilburn1994}, (\ref{SNRnoloss}) gives
\be
\delta \phi_{min}=1/N.\label{deltaphiminnoloss}
\ee
\xxx{deltaphiminnoloss}
\yyy{p. NSL-3 box1}

Phase measurement  by this method is thus seen to be at the Heisenberg limit (\ref{Heisenberglimit}), with a precision
that can be increased arbitrarily by increasing $N$\/.

\section{NOON-State Phase Measurement in the Presence of Loss}

In any real system  some photons will
inevitably be lost prior to detection, a feature not represented in the model of phase measurement 
described above.    Loss can be represented  by including in the model fictitious beam splitters\cite{GardinerZoller2004} through
which photons in the state  (\ref{NOONphistatedef}) pass before being subjected to the measurement (\ref{ANdef}).
Having in mind potential application to laser radar with coherent detection\cite{Kingston1995}, where one beam
impinges directly on  a detector while the other first suffers loss due to spreading during  reflection from
a distant target, we include a single such fictitious beam splitter, in mode $b$\/.

\begin{center}
\begin{figure}[t]
%\vspace*{2in}
\hspace*{1.35in}
%\special{wmf: testpic2.emf x=3.2in y=1.15in}
%\special{ps: testpic2.ps x=3.2in y=1.15in}
\includegraphics[width=3.2in,height=1.15in]{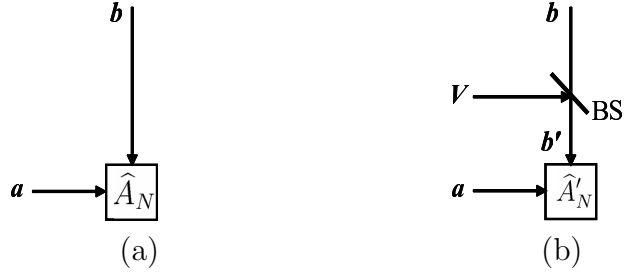}
\begin{center}
(a)\hspace*{2in}(b)
\end{center}
\caption{Input modes to phase measurement.  (a) Without loss.  (b) With loss in one mode, modeled by beam splitter BS.}
\label{Fig1}
\end{figure}
\end{center}

Denote  by $\wh{a}_b$\/ the
mode operator at the input port to the fictitious beam splitter, by $\wh{a}_{b\pr}$\/ the
mode operator at the output port of the beam splitter through which photons proceed to the detector, and by $\wh{a}_V$\/ the mode operator at the
other input port (vacuum port) of the beam splitter (see Fig.~\ref{Fig1}). 
These operators are related by\cite{GottfriedYan2003}
\be
\wh{a}_{b\pr}=t\wh{a}_b+r\wh{a}_V\label{bsops}
\ee
\xxx{bsops}
with $t$\/ and $r$\/ the respective transmission and reflection coefficients.
The loss  $L$\/ which is thus represented is of magnitude 
\be
L=1-\eta, \label{Ldef}
\ee
\xxx{Ldef}
where 
\be
\eta=|t|^2.\label{etadef}
\ee
\xxx{etadef}

The  detection operator (\ref{ANdef}) becomes
\be
\wh{A}\pr_N=|0,N\ra_{a,b\pr}\;\;\mbox{}_{a,b\pr}\la N,0|+|N,0\ra_{a,b\pr}\;\;\mbox{}_{a,b\pr}\la 0,N|.\label{ANprdef}
\ee
\xxx{ANdefpr}
\yyy{p. NSL-3 box 3}
The $a$\/ mode is unaffected by the presence of the beam splitter, and 
\be
|0\ra_b\/=|0\ra_{b\pr},\label{bvaceqbprvac} 
\ee
\xxx{bvaceqbprvac} 
since the beam splitter does not introduce additional photons into the system. Using (\ref{prodstatedef}), (\ref{numberstatedef}),
(\ref{bsops}), (\ref{ANprdef}) and (\ref{bvaceqbprvac}),
\bea
\wh{A}_N\pr=
\frac{1}{\sqrt{N}}\left[\left(t^\ast \wh{a}\da_b+r^\ast\wh{a}\da_V\right)^N\left(|0,0\ra_{a,b}\;\;\mbox{}_{a,b}\la N,0|\right) +
 \left(|N,0\ra_{a,b}\;\;\mbox{}_{a,b}\la 0,0|\right)\left(t\wh{a}_b+r\wh{a}_V\right)^N\right].\label{ANpr}
\eea
\xxx{ANpr}
\yyy{p. NSL-4, box 1}
The state space is now enlarged to include the fictitious beam splitter vacuum port mode $V$\/, so 
the state vector must include a factor of the vacuum state for that mode:
\be
|N::0;\phi\ra_{a,b,V}=|N::0;\phi\ra_{a,b}|0\ra_V.\label{NOONphistateV}
\ee
\xxx{NOONphistateV}
Using (\ref{NOONphistatedef}), (\ref{etadef}),  (\ref{ANpr}) and (\ref{NOONphistateV}), and defining
\be
\theta_t=\arg t,\label{thetatdef}
\ee
\xxx{thetatdef}
we obtain
\bea
\la \wh{A}\pr_N\ra_{\phi}&=&\mbox{}_{a,b,V}\la N::0;\phi|\wh{A}\pr_N|N::0;\phi\ra_{a,b,V}\nonumber\\
&=&\eta^{N/2}\cos(N(\phi +\theta_t))\label{ANprexpval}
\eea
\xxx{ANprexpval}
and
\bea
\mbox{Var}_{\phi}\wh{A}\pr_N&=&\mbox{}_{a,b,V}\la N::0;\phi|\left(\wh{A}\pr_N\right)^2|N::0;\phi\ra_{a,b,V}-\left(\mbox{}_{a,b,V}\la N::0;\phi|\wh{A}\pr_N|N::0;\phi\ra_{a,b,V}\right)^2\nonumber\\
&=&\frac{1}{2}\left(1+\eta^N\right)-\eta^N\cos^2\left(N(\phi + \theta_t)\right).\label{VarphiAprN}
\eea
\xxx{VarphiAprN}
\yyy{p. NSL-7, last box}
The signal-to-noise ratio for detecting a small change of phase $\delta \phi$ in the presence of loss, is, using (\ref{ANprexpval}) and 
(\ref{VarphiAprN}),
\bea
\mbox{SNR}\pr&=&\left(\la \wh{A}\pr_N\ra_{\phi_0+\delta\phi}-\la \wh{A}\pr_N\ra_{\phi_0}\right)^2/\mbox{Var}_{\phi_0}\wh{A}\pr_N\nonumber\\
&=&\frac{N^2\sin^2(N(\phi_0+\theta_t))\left(\delta \phi\right)^2}{\frac{1}{2}\left(\eta^{-N}+1\right)-\cos^2(N(\phi_0+\theta_t))}.\label{SNRpr}
\eea
\xxx{SNRpr}
\yyy{p. NSL-8, box 4, modified}

The minimum detectable phase change in the presence of loss,   that value of $\delta \phi$\/ for which $\mbox{SNR}\pr$\/ in (\ref{SNRpr})
is unity, is therefore
\be
\delta \phi\pr_{min}=
\frac{  \left[\frac{1}{2}\left(\eta^{-N}+1\right)-\cos^2(N(\phi_0+\theta_t))\right]^{1/2}  }{N|\sin(N((\phi_0+\theta_t))|}.
\label{deltaphiprmin}
\ee
\xxx{deltaphiprmin}
\yyy{p. NSL-19, box 2 , modified}
In the absence of loss, i.e. for $\eta=1$\/, (\ref{deltaphiprmin}) agrees with (\ref{deltaphiminnoloss}). 
For fixed $\eta < 1$\/ and $N$\/,
(\ref{deltaphiprmin}) is minimized for values of $\phi_0+\theta_t$\/ such that
\be
N(\phi_0+\theta_t)=(n+1/2)\pi,\hspace*{5mm}n=0,\pm 1,\pm2, \ldots  \label{minphasecond}
\ee
\xxx{minphasecond}
\yyy{p. NSL-20, box 2}
Since we wish to model pure loss we will take the transmission coefficient of the fictitious beam splitter to be real, so 
\be
\theta_t=0.\label{treal}
\ee
\xxx{treal}
Imposing (\ref{treal}) and assuming that $\phi_0$\/ satisfies (\ref{minphasecond}), (\ref{deltaphiprmin}) becomes
\be
\delta \phi\pr_{min}=\frac{\sqrt{\left(\eta^{-N}+1\right)/2}}{N}.
\label{deltaphiprmin2}
\ee
\xxx{deltaphiprmin2}
\yyy{p. NSL-21, box 2}
This result agrees with that obtained previously by Chen {\em et al.}\/ \cite{ChenJiangHan2006} using a master-equation model of continuous 
loss and entanglement.\footnote{The  model of \cite{ChenJiangHan2006} corresponds to that of the present paper 
when the parameters $\bar{\gamma} t$\/, $\Gamma_1 t$\/ and $\Gamma_2 t$\/ of the former are set to values of 0, 0 and $-\log\eta$\/,
respectively.}
For any nonzero amount of loss, i.e., for $\eta < 1$\/, we see from (\ref{deltaphiprmin2})  that 
\be
\lim_{N \rightarrow \infty}\delta \phi\pr_{min}=\infty.\label{limdeltaphiprmin}  
\ee
\xxx{limdeltaphiprmin}

\section{The Small-Loss and Large-Loss Cases}\label{Sec_small_loss}

The behavior of $\delta \phi\pr_{min}$\/ for varying $N$\/ and $\eta$\/, as given exactly in (\ref{deltaphiprmin2}), is of particular
interest in two limiting cases: very large amounts of loss, $L \ltsimeq 1$\/,  $\eta \ll 1$\/, and very small amounts of loss,
$L \ll 1$\/, $\eta \ltsimeq 1$\/. The large-loss limit is relevant for laser radar, 
while the limit of small loss, on the other hand, is relevant for precision laboratory experiments and technological
applications. 

\begin{center}
\begin{figure}[t]
%\vspace*{2.5in}
%\hspace*{.22in}
%\special{wmf: graphpic_a2.emf x=2.75in y=2.10in}
%\special{ps: graphpic_a2.ps x=2.75in y=2.10in}
\includegraphics[width=2.75in,height=2.10in]{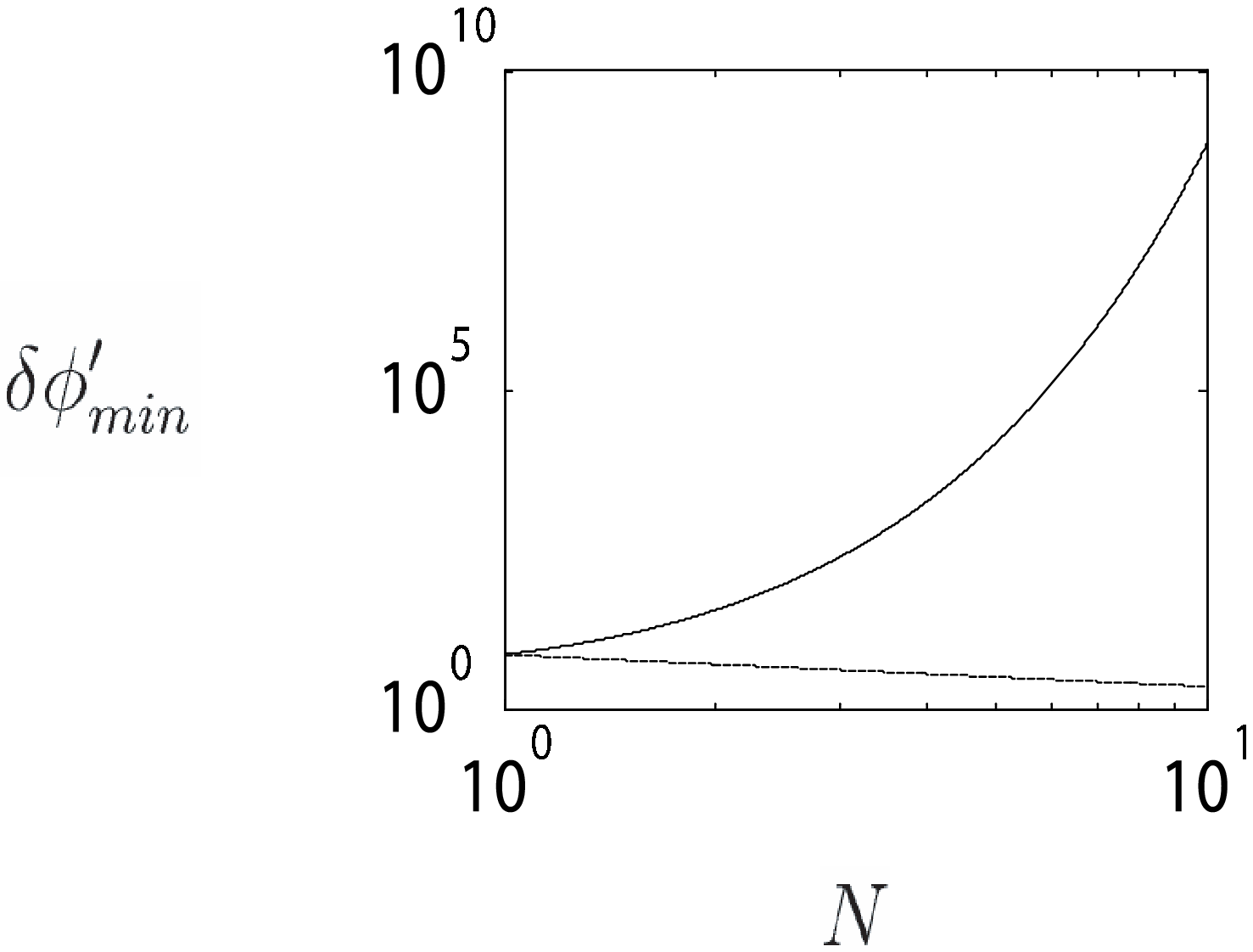}
\hspace*{.5in}
%\special{wmf: graphpic_b2.emf x=2in y=2.10in}
%\special{ps: graphpic_b2.ps x=2in y=2.10in}
\includegraphics[width=2in,height=2.10in]{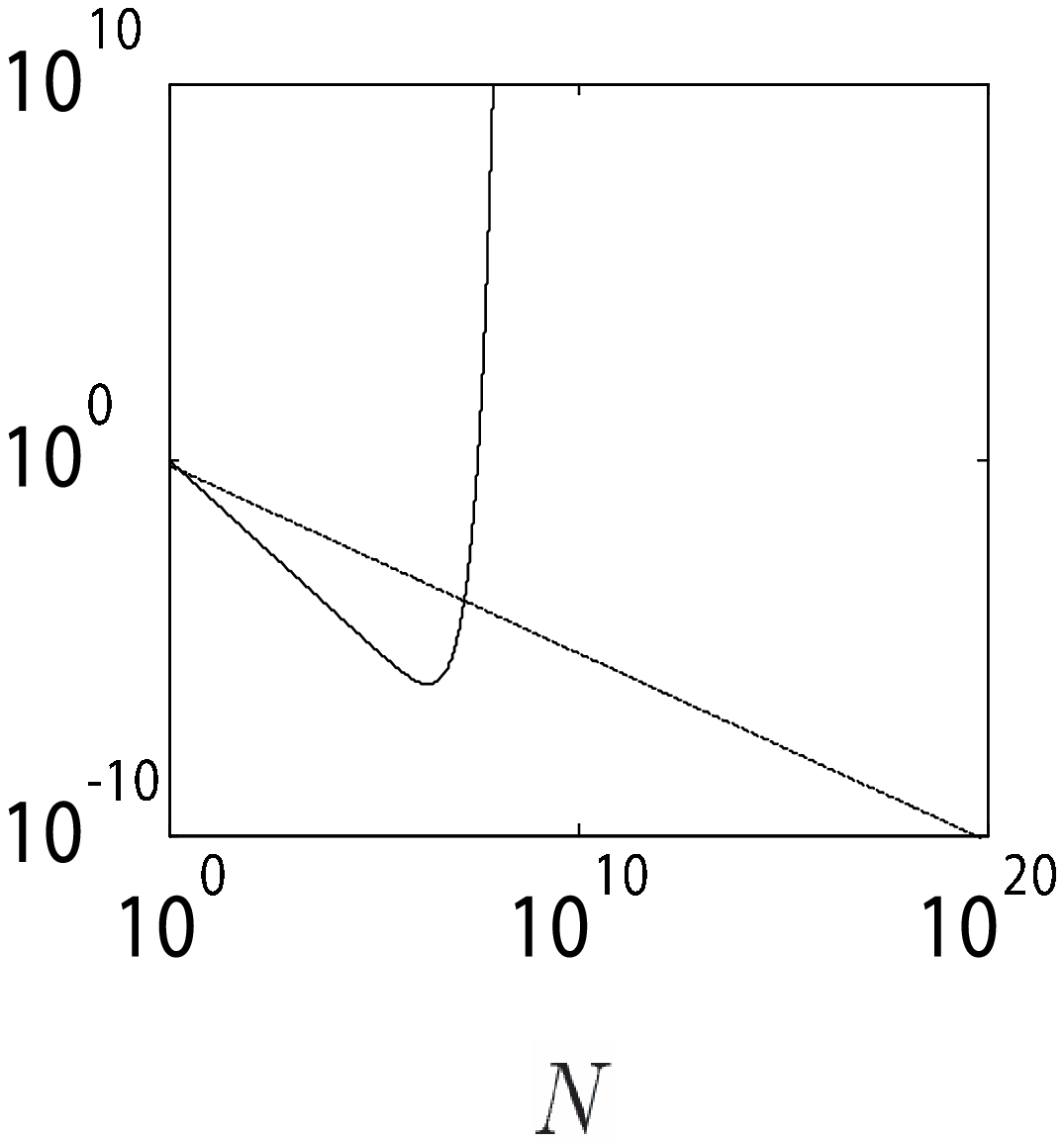}
\begin{center}
(a)\hspace*{2in}(b)
\end{center}
\caption{$\delta\phi\pr_{min}$\/ (curved lines) as a function of $N$\/.
(a) $L=.99$\/. (b) $L=10^{-6}$\/. Straight lines are the function $1/\sqrt{2\eta N}$\/, where $\eta=1-L.$ }
\label{Fig2}
\end{figure}
\end{center}

Consider first the case of large loss. From (\ref{deltaphiprmin2}),
\be
 \frac{d\delta \phi\pr_{min}}{dN} = -\frac{1}{N^2}\left[\left(\eta^{-N}+1\right)/2\right]^{1/2}
 -\frac{\eta^{-N}\log \eta}{4N}\left[\left(\eta^{-N}+1\right)/2\right]^{-1/2},\label{dphimprmindN}
\ee
\xxx{dphimprmindN}
\yyy{p. NSL-39, last box}
so
\be
\lim_{\eta \rightarrow 0} \frac{d\delta \phi\pr_{min}}{dN} =\frac{-\log \eta \; \eta^{N/2}}{4\sqrt{2}N}\label{limeta0dphimprmindN}
\ee
\xxx{limeta0dphimprmindN}
\yyy{p. NSL-42, last box}
which for $\eta < 1$\/ is strictly positive for all  $N$\/. So increasing $N$\/ can only harm the precision of phase measurement
in this limit, and there is no $N$\/ for which the detector can provide useful results satisfying 
(\ref{smallphasechange}). See Fig.~\ref{Fig2}(a).

In the limit of small loss, as exemplified in Fig. \ref{Fig2}(b), we can estimate the smallest possible $\delta \phi\pr_{min}$\/, and the value of $N$\/ at which
it is obtained, as follows. From (\ref{deltaphiprmin2}),
\be
\frac{d}{dN}\log\delta\phi\pr_{min}=-\frac{1}{N}-\frac{\log\eta}{2\left(\eta^N+1\right)},\label{ddNdeltaphiprmin}
\ee
\xxx{ddNdeltaphiprmin}
\yyy{p. NSL-43, box }
so
\be
N_{min}(\eta)=\frac{-2\left(\eta^{N_{min}(\eta)}+1\right)}{\log \eta},\label{Nmineq}
\ee
\xxx{Nmineq}
where $N_{min}(\eta)$\/ is that $N$\/ which minimizes $\delta \phi\pr_{min}$\/ for a given $\eta$\/.
We look for $N_{min}(\eta)$\/ of the form
\be
\lim_{L \rightarrow 0} N_{min}(\eta)=\frac{\nu}{L}.\label{Nminansatz}
\ee
\xxx{Nminansatz}
\yyy{p. NSL-43, box 6}
Using (\ref{Nminansatz}) in (\ref{Nmineq}), we obtain
\be
\frac{\nu}{L}=\lim_{L \rightarrow 0}\frac{-2\left( (1-L)^{\frac{\nu}{L}}+1\right)}{-L},\label{nueq1}
\ee
\xxx{nueq1}
\yyy{p. NSL-43, last box}
or
\be
\nu=2\left(e^{-\nu}+1\right)\label{nueq2}
\ee
\xxx{nueq2}
\yyy{p. NSL-44, box 1}
which may be solved numerically to obtain
\be
\nu\approx 2.218.\label{nuval}
\ee
\xxx{nuval}
\yyy{p. NSL-44, box 3}
Using (\ref{Nminansatz}) in (\ref{deltaphiprmin2})
\be
\lim_{L \rightarrow 0} \delta\phi\pr_{min}\;\rule[-2mm]{.1mm}{7mm}_{N=N_{min}(\eta)}=\mu L,\label{deltaphiprminmin}
\ee
\xxx{deltaphiprminmin}
\yyy{p. NSL-44, last box}
where
\bea
\mu&=&\lim_{L \rightarrow 0} \frac{1}{\nu}\left[\frac{1}{2}\left( (1-L)^{-\frac{\nu}{L}}+1\right)\right]^{1/2}\nonumber\\
&=&\frac{1}{\nu}\left[\frac{1}{2}\left(e^\nu+1\right)\right]^{1/2}\nonumber\\
&\approx&1.018. \label{muval}
\eea
\xxx{muval}
\yyy{p. NSL-44, last box, modified}
using (\ref{nuval}).
For $L$\/ as large as .01 the expressions (\ref{Nminansatz}) and (\ref{deltaphiprminmin}) give values within a percent of the exact values obtained from (\ref{deltaphiprmin2}).   

To find the critical value of  loss $L=L_c$\/ above which  $\delta\phi\pr_{min}$\/  must be a nondecreasing function of $N$\/, we first examine the cases $N=1$\/ and $N=2$\/.
For $\delta \phi\pr_{min}$\/ to be smaller at $N=2$\/ than at $N=1$\/, we find  from (\ref{deltaphiprmin2}) that we must have
\be
\eta > \eta_c,
\ee
\xxx{etacrit}
\yyy{p. NSL-41, last box}
where
\be
\eta_c = \frac{\sqrt{7}-2}{3}\approx 0.215.\label{etacritval}
\ee
\xxx{etacritval}
\yyy{p. NSL-41, last box}
From this it follows that, if $\eta \le \eta_c$\/, then $\delta \phi\pr_{min}$\/ will not be smaller than its value at $N=2$\/ for
any value of $N$\/. For, if $\delta \phi\pr_{min}$\/ were to be smaller for some $N >2$\/ than for $N=2$\/,  
it would be necessary for
\be
\frac{d\delta \phi\pr_{min}}{dN} < 0 \label{ddeltaphiprminlt0}
\ee
\xxx{ddeltaphiprminlt0}
\yyy{p. NSL-54}
to hold for some value of $N \ge 2$. Using (\ref{dphimprmindN}), this means that for some $N \ge 2$,
\be
\log \eta > \frac{-2}{N}\left(\eta^N+1\right). \label{logetacondition}
\ee
\xxx{logetacondition}
\yyy{p. NSL-54 last box}
But  $\eta \le \eta_c$\/, so (\ref{logetacondition}) implies
\be
\log \eta > \frac{-2}{N}\left({\eta_c}^N+1\right) \label{logetacondition2}
\ee
\xxx{logetacondition2}
\yyy{p. NSL-55, 2nd box}
and
\be
\log \eta_c > \frac{-2}{N}\left({\eta_c}^N+1\right). \label{logetacondition3}
\ee
\xxx{logetacondition3}
\yyy{p. NSL-55, 3nd box}
So,
\be
N < \frac{-2\left({\eta_c}^N+1\right)}{\log \eta_c}, \label{Ncond}
\ee
\xxx{Ncond}
\yyy{p. NSL-55}
implying
\be
N < \frac{-2\left({\eta_c}+1\right)}{\log \eta_c} \label{Ncond2}
\ee
\xxx{Ncond2}
\yyy{p. NSL-55, 4th box}
since $\eta_c < 1$ and  $N \ge 2$. Using (\ref{Ncond2}) and (\ref{etacritval}), we obtain
\be
N \ltsimeq 1.582, \label{Ncontradiction}
\ee
\xxx{Ncontradiction}
\yyy{p. NSL-55, 5th box}
contradicting the requirement $N \ge 2$\/. So $\delta\phi\pr_{min}$\/ will be a nondecreasing function
of $N$\/ whenever
\be
L > L_c,\label{Lcrit}
\ee
\xxx{Lcrit}
where
\be
L_c =1-\eta_c \approx 0.785.\label{Lcrit2}
\ee
\xxx{Lcrit2}

\section{Comparison With Unentangled Phase Measurement; Multiple Measurements}

For phase estimation with unentangled coherent light and homodyne or heterodyne de\-tec\-tion\cite{GardinerZoller2004}, we would expect a precision of $\kappa/\sqrt{\eta N}$\/ in the presence of
loss, where $\kappa$\/ is independent of $N$\/ and of order unity. No matter how large the loss, this precision can always be improved by increasing $N$\/, and thus can
always surpass the precision attainable with NOON states and a detector implementing the operator  
(\ref{ANdef}). (It is conceivable that detectors implementing other measurement operators, with nonvanishing matrix elements between states other than just linear combinations of $|N,0\ra_{a,b}$\/
and $|0,N\ra_{a,b}$\/, might be less sensitive to loss while still surpassing the standard quantum limit (\ref{SQL}), but we have not investigated this issue here.)    If in a particular application with small loss
there is a limit to how large $N$\/ can be, and if this limit is not much larger
than that given by (\ref{Nminansatz}), (\ref{nuval}),  then the use of NOON states with (\ref{ANdef})
can lead to  precision better than that attainable with standard techniques. See Fig.~{\ref{Fig2}(b).

The  analysis up to this point has been based on phase measurements using individual quantum states with $N$\/ photons. If the measurements are repeated $M$\/ times,  using $M$\/ independent quantum states, the minimum detectable
phase change will decrease by an additional factor of $1/\sqrt{M}$\/. For measurement with unentangled coherent-state photons, the precision will be 
\be
\delta\phi_{un}=\kappa/\sqrt{\eta N M}=\kappa/\sqrt{\eta N_T}, \label{deltaphiun}
\ee
\xxx{deltaphiun}
where 
\be
N_T=NM \label{NTdef}
\ee 
\xxx{NTdef}
is the average total number of photons  available. That is, for phase measurements with unentangled coherent light we obtain the same
precision whether we make many measurements with fewer photons per measurement or fewer measurements with more photons per measurement. 

For NOON-state photons, the precision after $M$\/ $N$\/-photon measurements is
\bea
\delta\phi_{NOON}&=&\delta \phi\pr_{min}/\sqrt{M}\nonumber\\
                 &=&\sqrt{\frac{\eta^{-N}+1}{2NN_T}}\label{deltaphiNOON}
\eea
\xxx{deltaphiNOON}
using (\ref{deltaphiprmin2}) and (\ref{NTdef}), or
\be
\delta\phi_{NOON}=R_{NOON}/\sqrt{\eta N_T},\label{deltaphiNOON2}
\ee
\xxx{deltaphiNOON2}
where $R_{NOON}$ is, aside from the constant factor $\kappa$\/, the ratio of NOON phase measurement precision to unentangled phase precision (\ref{deltaphiun}) with equal $L$\/ and $N_T$\/,
\be
R_{NOON}=\sqrt{\frac{\eta(\eta^{-N}+1)}{2N}}.\label{RNOON}
\ee
\xxx{RNOON}
Graphs of $R_{NOON}$\/ as a function of $N$\/ are presented in Fig. \ref{Fig3} for $L=.99$\/ and $L=10^{-6}$\/. 

For fixed $N_T$\/, $\delta\phi_{un}$\/ is constant, and $\delta\phi_{NOON}$\/
is minimized by minimizing $R_{NOON}$\/ as a function of  the number $N$\/ of photons per NOON state. 
Denote the minimizing value of $N$\/  by $\wti{N}_{min}(\eta)$\/. 
For large loss, $L\ltsimeq 1$\/, we find from (\ref{RNOON}) that
\be
\lim_{\eta\rightarrow 0}\frac{dR_{NOON}}{dN}= \frac{-\log \eta \; \eta^{-\left(\frac{N-1}{2}\right)}}{8\sqrt{2N}}
\label{dRdNlargeloss}
\ee
\xxx{dRdNlargeloss}
\yyy{p. NSL-61, top box}
which is strictly positive for all $N$\/. (See, e.g., Fig. \ref{Fig3}(a).)  So
\be
\lim_{L\rightarrow 1}\wti{N}_{min}(\eta)=1,\label{NtiminL1}
\ee
\xxx{NtiminL1}
which with (\ref{deltaphiNOON}) yields
\be
\lim_{L \rightarrow 1} \delta\phi_{NOON}\;\rule[-2mm]{.1mm}{7mm}_{\;N=\wti{N}_{min}(\eta)}=1/\sqrt{2\eta N_T}.
\label{deltaphiNOONminL1}
\ee
\xxx{deltaphiNOONminL1}
The phase measurement precision obtainable with NOON states is thus, in the large-loss limit, % $L\approx 1$\/,
the same  as that obtainable with unentangled coherent states, eq. (\ref{deltaphiun}), up to a constant factor.

In the complete absence of loss, i.e. for $\eta=1$\/, $R_{NOON}=1/\sqrt{N}$\/ and is minimized by making $N$\/ as large as possible, 
\be
\wti{N}_{min}(\eta)\;\rule[-2mm]{.1mm}{7mm}_{\; L=0}=N_T. \label{NtiminLzero}
\ee
\xxx{NtiminLzero}
That is, for $L=0$\/ the greatest precision using the NOON-state measurement scheme (\ref{ANdef}) and a fixed
total number of photons $N_T$\/ is obtained by making a single measurement with all $N_T$\/ photons.  
Using (\ref{NtiminLzero}) and $\eta=1$\/ in (\ref{deltaphiNOON}),
\be
\delta\phi_{NOON}\;\rule[-2mm]{.1mm}{7mm}_{\;N=N_{min}(\eta), L=0}=1/N_T.\label{deltaphiNOONLzero}
\ee
\xxx{deltaphiNOONLzero}
Using $\eta=1$\/ in (\ref{deltaphiun}),
\be
\delta\phi_{un}\;\rule[-2mm]{.1mm}{7mm}_{\;L=0}=\kappa/\sqrt{N_T}.\label{deltaphiunLzero}
\ee
\xxx{deltaphiunLzero}
Comparing (\ref{deltaphiNOONLzero}) and (\ref{deltaphiunLzero}) we see that, in the absence of loss, the improvement in phase measurement precision obtained by
using NOON states is of order $\sqrt{N_T}$\/, as expected.

For small loss, $\eta \ltsimeq 1$\/, 
an analysis along the lines of Sec. \ref{Sec_small_loss} gives
\be
\lim_{L \rightarrow 0} \wti{N}_{min}(\eta)=\frac{\wti{\nu}}{L},\label{Ntiminansatz}
\ee
\xxx{Ntiminansatz}
where $\wti{\nu}$\/ is the solution to
\be
\wti{\nu}=e^{-\wti{\nu}}+1\label{nutieq2},
\ee
\xxx{nutieq2}
which is found numerically to be
\be
\wti{\nu}\approx 1.279.\label{nutival}
\ee
\xxx{nutival}
\yyy{p. NSL-58}
The corresponding minimum value of $\delta\phi_{NOON}$\/ is
\be
\lim_{L \rightarrow 0} \delta\phi_{NOON}\;\rule[-2mm]{.1mm}{7mm}_{\;N=\wti{N}_{min}(\eta)}=\wti{\mu}\sqrt{ L/N_T},
\label{deltaphiNOONmin}
\ee
\xxx{deltaphiNOONmin}
where
\be
\wti{\mu}=\sqrt\frac{e^{\wti{\nu}}+1}{2\wti{\nu}}\approx 1.340.\label{mutival}
\ee
\xxx{mutival}
\yyy{p. NSL-59}
Comparing ({\ref{deltaphiNOONmin}) with (\ref{deltaphiun}), we see that, when $L\gtsimeq 0$\/, NOON states give
an improvement in phase measurement precision of order $\sqrt{L}$\/.

(In the limit of zero loss, (\ref{Ntiminansatz}) indicates that $R_{NOON}$\/ has  a local minimum
at $\wti{N}_{min}(1)=\infty$\/, corresponding according to (\ref{deltaphiNOONmin}) to $\delta\phi_{NOON}=0$\/. But, of course,
$N$\/ cannot be made larger than $N_T$\/, corresponding to the results (\ref{NtiminLzero}), (\ref{deltaphiNOONLzero})
in the lossless case.)

An analysis along the lines of  Sec. \ref{Sec_small_loss} shows that $R_{NOON}$\/, and therefore
$\delta\phi_{NOON}$\/ for fixed $N_T$\/, 
is an increasing function of $N$\/ for all $L > \wti{L}_c$\/, where 
\be
\wti{L}_c=2- \sqrt{2} \approx 0.586.\label{Ltic}
\ee
\xxx{Ltic}
\yyy{pp. NSL-61-65}
It is not surprising that $\wti{L}_c$\/ is lower
than $L_c$\/ since, in the multiple-measurement case, increasing $N$\/, even when it
decreases the single-measurement precision $\delta\phi\pr_{min}$\/, increases the
factor $1/\sqrt{M}=\sqrt{N_T/N}$\/ which enters into $\delta\phi_{NOON}$\/, eq. (\ref{deltaphiNOON}).

\begin{center}
\begin{figure}[t]
%\vspace*{2.5in}
%\hspace*{.2in}
%\special{wmf: fig3a.emf x=2.75in y=2.10in}
%\special{ps: fig3a.ps x=2.75in y=2.10in}
\includegraphics[width=2.75in,height=2.10in]{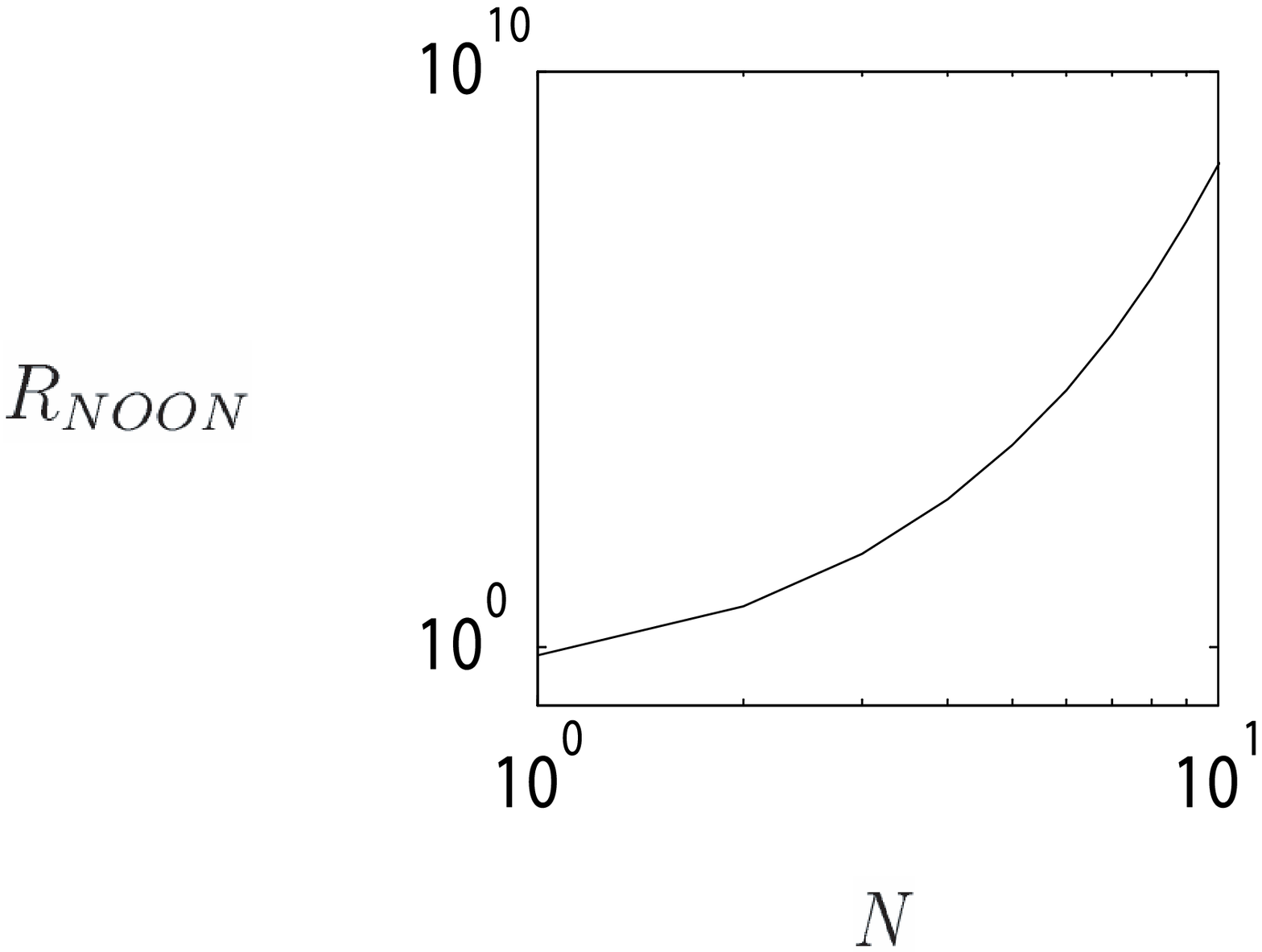}
\hspace*{.5in}
%\special{wmf: fig3b.emf x=1.97in y=2.10in}
%\special{ps: fig3b.ps x=1.97in y=2.10in}
\includegraphics[width=1.97in,height=2.10in]{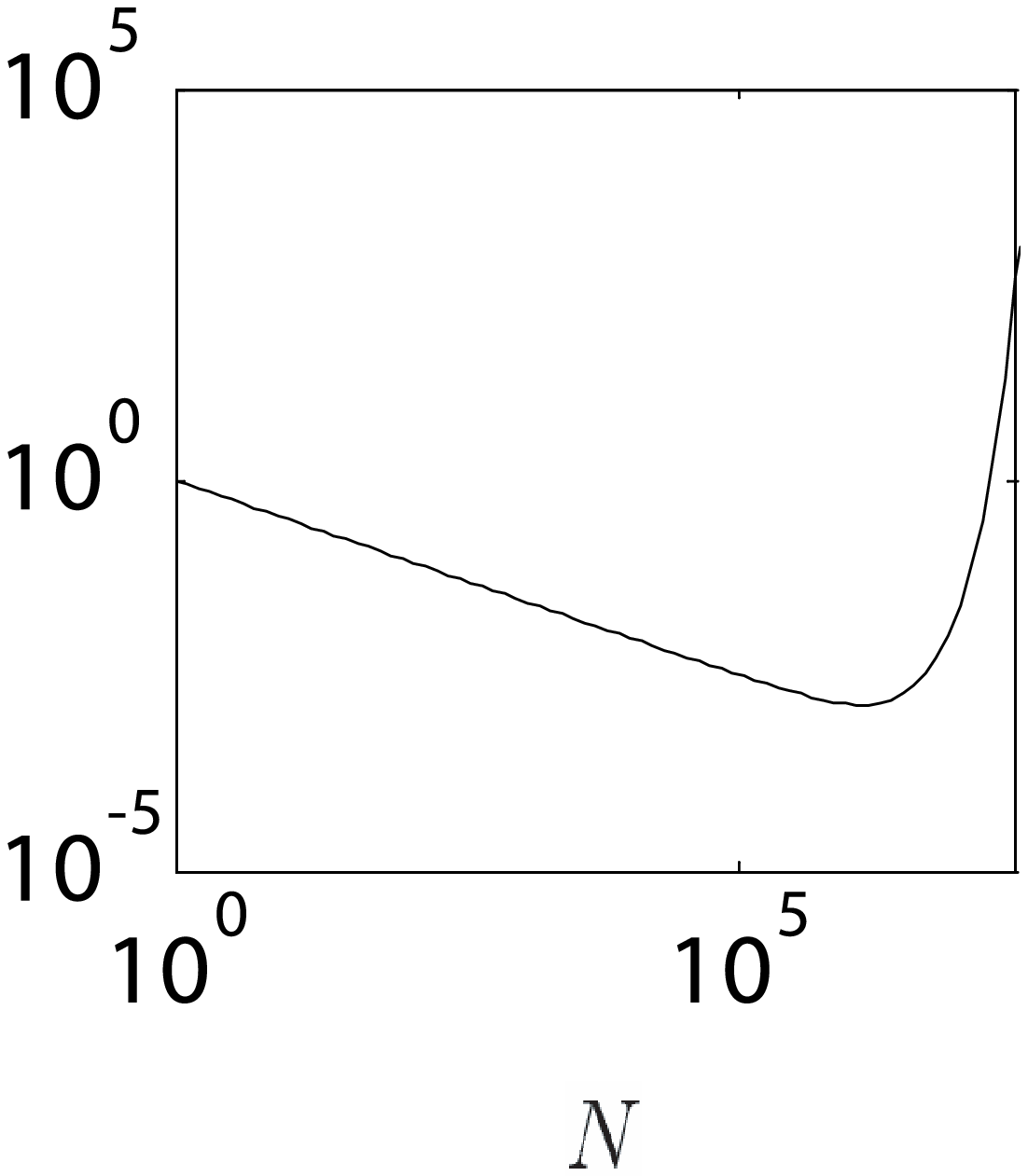}
\begin{center}
(a)\hspace*{2in}(b)
\end{center}
\caption{$R_{NOON}$\/ as a function of $N$\/. (a) $L=.99$\/. (b) $L=10^{-6}$\/.}
\label{Fig3}
\end{figure}
\end{center}

\end{document}